\documentclass[useAMS,usenatbib]{mn2e}

\usepackage{graphicx}	
\usepackage{amsmath}	
\usepackage{amssymb}	
\usepackage{upgreek}
\newcommand{\MS}{\ifmmode{\,}\else\thinspace\fi{\rm M}\ifmmode_{\odot}\else$_{\odot}$\fi}
\newcommand{\LS}{\ifmmode{\,}\else\thinspace\fi{\rm L}\ifmmode_{\odot}\else$_{\odot}$\fi}
\newcommand{\RS}{\ifmmode{\,}\else\thinspace\fi{\rm R}\ifmmode_{\odot}\else$_{\odot}$\fi}
\newcommand{\ff}{\ifmmode \nu_{\rm F}\else$\nu_{\rm F}$\fi}
\newcommand{\fo}{\ifmmode \nu_{\rm 1O}\else$\nu_{\rm 1O}$\fi}
\newcommand{\fx}{\ifmmode \nu_{\rm x}\else$\nu_{\rm x}$\fi}
\newcommand{\ax}{\ifmmode A_{\rm x}\else$A_{\rm x}$\fi}
\newcommand{\Po}{\ifmmode P_{\rm 1O}\else$P_{\rm 1O}$\fi}
\newcommand{\Pf}{\ifmmode P_{\rm F}\else$P_{\rm F}$\fi}
\newcommand{\Px}{\ifmmode P_{\rm x}\else$P_{\rm x}$\fi}
\newcommand{\pxpo}{\ifmmode P_{\rm x}/P_{\rm 1O}\else$P_{\rm x}/P_{\rm 1O}$\fi}
\newcommand{\pxpf}{\ifmmode P_{\rm x}/P_{\rm F}\else$P_{\rm x}/P_{\rm F}$\fi}

\title[Peculiar double-periodic pulsation in RR~Lyr stars]{Peculiar double-periodic pulsation in RR~Lyrae stars of the OGLE collection. I. Long-period stars with dominant radial fundamental mode.}
\author[Smolec et al.]{
R. Smolec$^{1}$\thanks{E-mail: smolec@camk.edu.pl}, Z. Prudil$^{2}$, M. Skarka$^{3}$, K. B\c{a}kowska$^{1,4}$
\\
$^{1}$ Nicolaus Copernicus Astronomical Center, Polish Academy of Sciences, ul. Bartycka 18, 00-716 Warszawa, Poland\\
$^{2}$ Department of Theoretical Physics and Astrophysics, Masaryk University, Kotl\'a\v{r}sk\'a 2, 611 37 Brno, Czech Republic\\
$^{3}$ Konkoly Observatory, MTA Research Centre for Astronomy and Earth Sciences, Konkoly Thege Mikl\'{o}s \'{u}t 15-17, H--1121 Budapest, Hungary\\
$^{4}$ Fulbright Visiting Scholar, The Ohio State University, Dept. of Astronomy, 140 W. 18th Ave, Columbus, OH 43210, USA\\
}

\begin{document}

\date{Accepted . Received ; in original form }

\pagerange{\pageref{firstpage}--\pageref{lastpage}} \pubyear{2015}

\maketitle

\label{firstpage}

\begin{abstract}
We present the discovery of a new, peculiar form of double-periodic pulsation in RR~Lyrae stars. In four, long-period ($P>0.6$\thinspace d) stars observed by the Optical Gravitational Lensing Experiment, and classified as fundamental mode pulsators (RRab), we detect additional, low-amplitude variability, with period shorter than fundamental mode period. The period ratios fall in a range similar to double-mode fundamental and first overtone RR~Lyrae stars (RRd), with the exception of one star, in which the period ratio is significantly lower and nearly exactly equals 0.7. Although period ratios are fairly different for the four stars, the light curve shapes corresponding to the dominant, fundamental mode are very similar. The peak-to-peak amplitudes and amplitude ratios (Fourier parameters $R_{21}$ and $R_{31}$) are among the highest observed in RRab stars of similar period, while Fourier phases ($\varphi_{21}$ and $\varphi_{31}$) are among the lowest observed in RRab stars. If the additional variability is interpreted as due to radial first overtone, then, the four stars are the most extreme RRd variables of the longest pulsation periods known. Indeed, the observed period ratios can be well modelled with high metallicity pulsation models. However, at such long pulsation periods, first overtone is typically damped. Five other candidates, with weak signature of additional variability, sharing the same characteristics, were also detected and are briefly discussed.
\end{abstract}


\section{Introduction}\label{sec:intro}
For decades, RR~Lyrae stars were considered simple, large-amplitude, purely radial pulsators. Only a few forms of pulsation were detected: pulsation in the radial fundamental mode (F mode, RRab stars), pulsation in the radial first overtone mode (1O mode, RRc stars) and simultaneous pulsation in these two modes (RRd stars). The only known complication was the Blazhko effect -- quasi-periodic modulation of pulsation amplitude and/or phase, observed in a significant fraction of RRab stars and less commonly in RRc stars \citep[for recent reviews see e.g.][]{szabobl,sm15bl}.

With the enormous amount of top-quality data gathered by photometric sky surveys, most notably by the Optical Gravitational Lensing Experiment \citep[OGLE;][]{ogleIII,ogleIV}, and ultra-precise photometry collected by space missions, {\it MOST}, {\it CoRoT} and {\it Kepler}, new classes of double-mode RR~Lyr stars were discovered. In the most numerous group, in addition to the dominant radial first overtone, shorter period variability is detected. Period ratios of the two modes are in the $P_{\rm x}/P_{\rm1O}\in(0.6,0.64)$ range. First star of this type was detected with {\it MOST} \citep[][]{aqleo}, several stars were detected with {\it Kepler} and {\it CoRoT} \citep[e.g.][]{pamsm15,szabo_corot}. However, the largest sample was detected in the OGLE Galactic bulge data \citep{netzel1,netzel3}. For the recent explanation of the nature of the additional mode see \cite{wd16}. Another group of double-periodic stars was also detected in the Galactic bulge OGLE data. The dominant variability is again due to radial first overtone. Additional periodicity is of longer period, even longer than the expected period of the fundamental mode in these stars; the period ratios tightly cluster around $P_{\rm 1O}/P_{\rm x}\approx 0.686$ \citep{netzel2,nspta}. At the moment we have no explanation for the nature of the additional variability in these stars. A single, peculiar triple-mode RR~Lyr star with period doubling was reported by \cite{rs15b}. Finally, Blazhko effect was discovered in RRd stars in the Galactic bulge \citep{ogleIV_rrl_blg,rs15a}, in the globular cluster M3 \citep{jurcsikM3} and recently in the OGLE Magellanic Clouds' photometry \citep{ogleIV_rrl_mc}. These stars however, have non-typical period ratios -- almost always systematically lower than observed for non-modulated RRd stars (with similar fundamental mode periods). A few RR~Lyr stars with low amplitude variability and period ratios close to expected in RRd stars were also detected in the {\it Kepler} data \citep{benko10,benko14,molnar12} -- for a recent review see \cite{molnarRRL15}.

In this paper we report the discovery of new and puzzling form of double-mode RR~Lyr pulsation. In four long-period stars ($P>0.6$\thinspace d) observed by the OGLE project and catalogued as RRab stars \citep{ogleIII_rrl_blg,ogleIV_rrl_blg,pietruk}, we detect additional variability of shorter period, $\Px$. OGLE-BLG-RRLYR-01136, OGLE-BLG-RRLYR-07283 and OGLE-BLG-RRLYR-09116 are members of the Galactic bulge, while OGLE-GD-RRLYR-0035 is a Galactic disc star. In the following, we will refer to these stars simply as 01136, 07283, 09116 and 0035, respectively. Period ratios for these four stars are in the $P_{\rm x}/P_{\rm F}\!\in\!(0.7,0.74)$ range, similar to the RRd stars. However, their fundamental mode periods are longer than in any RRd star known. Period ratios are also typically lower than in RRd stars. We present detailed analysis of the photometric data for these stars and of pulsation models that indicate, that in all four cases double-mode, radial, F+1O pulsation is the most likely interpretation. Thus, the four stars are extreme RRd variables with the longest periods known. A few additional candidates, showing the same form of pulsation, but in which additional variability is not certain, are also identified.

\section{Data analysis}\label{sec:methods}

Our analysis follows a standard consecutive prewhitening technique. Significant (${\rm S/N}>4.0$) periodicities identified in the data with the help of the discrete Fourier transform, are included in the sine series of the following form:
\begin{equation}
m(t)=A_0+\sum_k A_k\sin(2\uppi\nu_k t + \phi_k),
\label{eq:sineseries}
\end{equation}
which is fitted to the data using non-linear least-square algorithm. Amplitudes, $A_k$, phases, $\phi_k$, and frequencies, $\nu_k$, are all adjusted. Prewhitened data are again inspected for the presence of additional signals. In eq.~\ref{eq:sineseries} we include only resolved frequencies. We consider two peaks as well resolved if their separation is larger than $2/\Delta T$, where $\Delta T$ is data length. Residual data are inspected for the presence of slow trends which are removed with the help of low-order polynomials or spline functions. Outliers are removed from the data (4$\sigma$ clipping). 

Typical solution includes two independent frequencies, of the fundamental mode, $\ff$, and of the additional variability, $\fx$, harmonics of the fundamental mode, $k\ff$, and linear frequency combinations, typically $k\ff+\fx$. If available, both OGLE-II/III and OGLE-IV data are analysed. If needed, the data are merged \citep[after detrending and levelling off the possible zero-point differences, see Tab.~\ref{tab:tab} and fig.~1 in][]{ogleIV_rrl_blg}. We analyse $I$-band data only, as these are much more numerous than $V$-band data.

For all stars the magnitude data were also transformed to the flux data (of arbitrary mean level). Then, it was checked whether all periodicities detected in the magnitude data persist (are significant) in the flux data. It was the case for all four stars. We note that the presence of combination frequencies in the flux data is a proof that the two involved frequencies originate from the same star, and are not e.g. due to blending with other source. In all stars a few combination frequencies were detected and hence, we conclude, that the additional periodicities are intrinsic to the analysed stars. In principle, the combination frequencies may also arise for two independent signals, due to the non-linearity of the detector. Indeed, the detectors used during OGLE-IV do show a non-linear response, but it is very small and restricted to very bright targets ($\langle I\rangle\lesssim 13$\thinspace mag), close to the saturation limit \citep{skowron}. This is highly unlikely, that this might be an issue for stars we consider here, all fainter than 15\thinspace mag.

The four stars were found by us while we were analysing the OGLE data for other projects. Then, we decided to check all long period RRab stars, with $P>0.6$\thinspace d, for the presence of additional low-amplitude periodicities falling in a similar range of period ratios, $\pxpf\!\in\!(0.69,\,0.76)$. The analysis was semi-automatic. Frequency of the dominant fundamental mode was identified first. Then, the data were prewhitened with the Fourier series including the fundamental mode frequency and its first 12 harmonics. Instead of standard prewhitening, a time-dependent prewhitening on a season-to-season basis was used \citep{pamsm15}. This has the advantage of removing the trends, possibly present in the data, and non-stationary variation associated with the fundamental mode (e.g. period change or long-period Blazhko effect). As a result, the number of confusing daily aliases in the frequency spectrum is reduced. The relevant frequency range was then inspected for the presence of additional significant periodicities. Photometric data for the potentially interesting stars were analysed manually. A few interesting additional candidates were selected and are discussed in Sect.~\ref{ssec:candidates}.

\section{Results}\label{sec:results}

\subsection{Overview}

\begin{table*}
\caption{Basic data for the four double-periodic stars (sorted by the increasing period). The consecutive columns contain: star's id, data source, mean $I$-band brightness, period of the fundamental mode, $\Pf$, period of the additional mode, $\Px$, period ratio, $\pxpf$, amplitude of the fundamental mode, $A_{\rm F}$, and amplitude ratio, $A_{\rm x}/A_{\rm F}$.}
\label{tab:tab}
\begin{tabular}{lrrr@{.}lr@{.}lrrrr}
star's id & source & $\langle I\rangle$\thinspace (mag) & \multicolumn{2}{c}{$\Pf$\thinspace (d)} & \multicolumn{2}{c}{$\Px$\thinspace (d)} & $\pxpf$ & $A_{\rm F}$\thinspace (mag) & $A_{\rm x}/A_{\rm F}$ & \\
\hline
OGLE-BLG-RRLYR-01136 & OGLE-IV     & 16.081 & 0&64549886(9) & 0&470494(4) & 0.7289 & 0.2054(2) & 0.015 &  \\ 
                     & OGLE-III    & 16.084 & 0&6455087(5)  &  \multicolumn{2}{c}{--} &  --    & 0.2088(9) &  --    & \\ 
OGLE-BLG-RRLYR-09116 & OGLE-IV     & 16.315 & 0&7326040(2)  & 0&54217(6)  & 0.7401 & 0.2022(3) & 0.023 & \\ 
                     & OGLE-III    & 16.306 & 0&7325679(3)  &  \multicolumn{2}{c}{--} &  --    & 0.1967(7) &  --    & \\ 
OGLE-GD-RRLYR-0035   & OGLE-III    & 17.326 & 0&73367608(9) & 0&534563(3) & 0.7286 & 0.2094(4) & 0.021 &  \\ 
OGLE-BLG-RRLYR-07283 & OGLE-IV     & 15.603 & 0&9214683(1)  & 0&6450353(6)& 0.7000 & 0.2061(1) & 0.082 & \\ 
                     & OGLE-II/III & 15.594 & 0&9214687(1)  & 0&6450503(5)& 0.7000 & 0.2108(2) & 0.085 & \\ 
\hline
\end{tabular}
\end{table*}

\begin{figure}
\centering
\resizebox{\hsize}{!}{\includegraphics{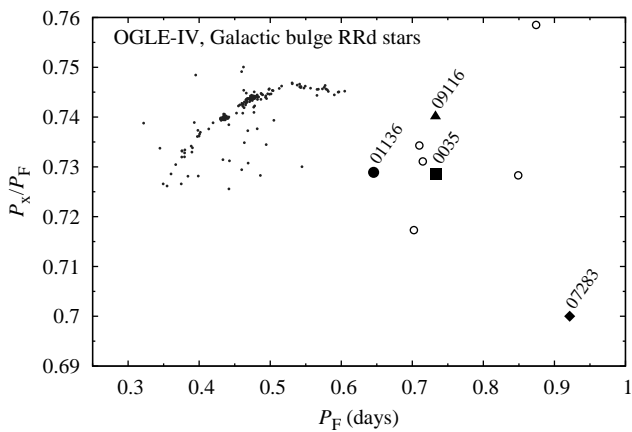}}
\caption{Petersen diagram for the discussed double-periodic stars (large symbols). Candidates are also included (open circles). For comparison, 173 OGLE-IV Galactic bulge RRd stars are plotted.}
\label{fig:pet}
\end{figure}

The basic parameters of the analysed stars (periods, period ratios and amplitudes of the detected periodicities) are collected in Tab.~\ref{tab:tab}. If both OGLE-III and OGLE-IV data were available, the two rows list the results of their separate analysis. The period ratios are plotted in the Petersen diagram in Fig.~\ref{fig:pet} and are compared with the period ratios of RRd stars detected in the Galactic bulge by the OGLE project \citep{ogleIV_rrl_blg}. The fundamental mode period in the four discussed stars is longer than in any other RRd star known so far; otherwise the period ratios of the considered stars, except of 07283, are in the range characteristic for the Galactic bulge RRd stars. For 07283, which is also of longest period in our small sample, it is significantly lower and nearly exactly equals to $0.7$, $P_{\rm x}/P_{\rm F}=0.7000081(7)$ (OGLE-IV).

\begin{figure}
\centering
\resizebox{\hsize}{!}{\includegraphics{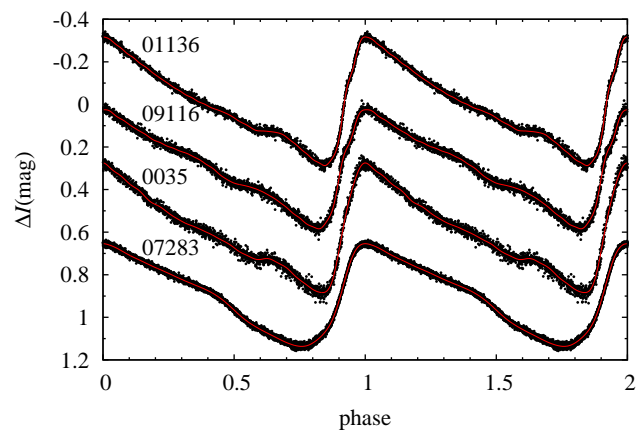}}
\caption{Light curves corresponding to the dominant fundamental mode for the four studied stars (after filtering out the additional variability). For clarity, light curves are shifted by $0.3$\thinspace mag with respect to each other.}
\label{fig:lc}
\end{figure}

\begin{figure}
\centering
\resizebox{\hsize}{!}{\includegraphics{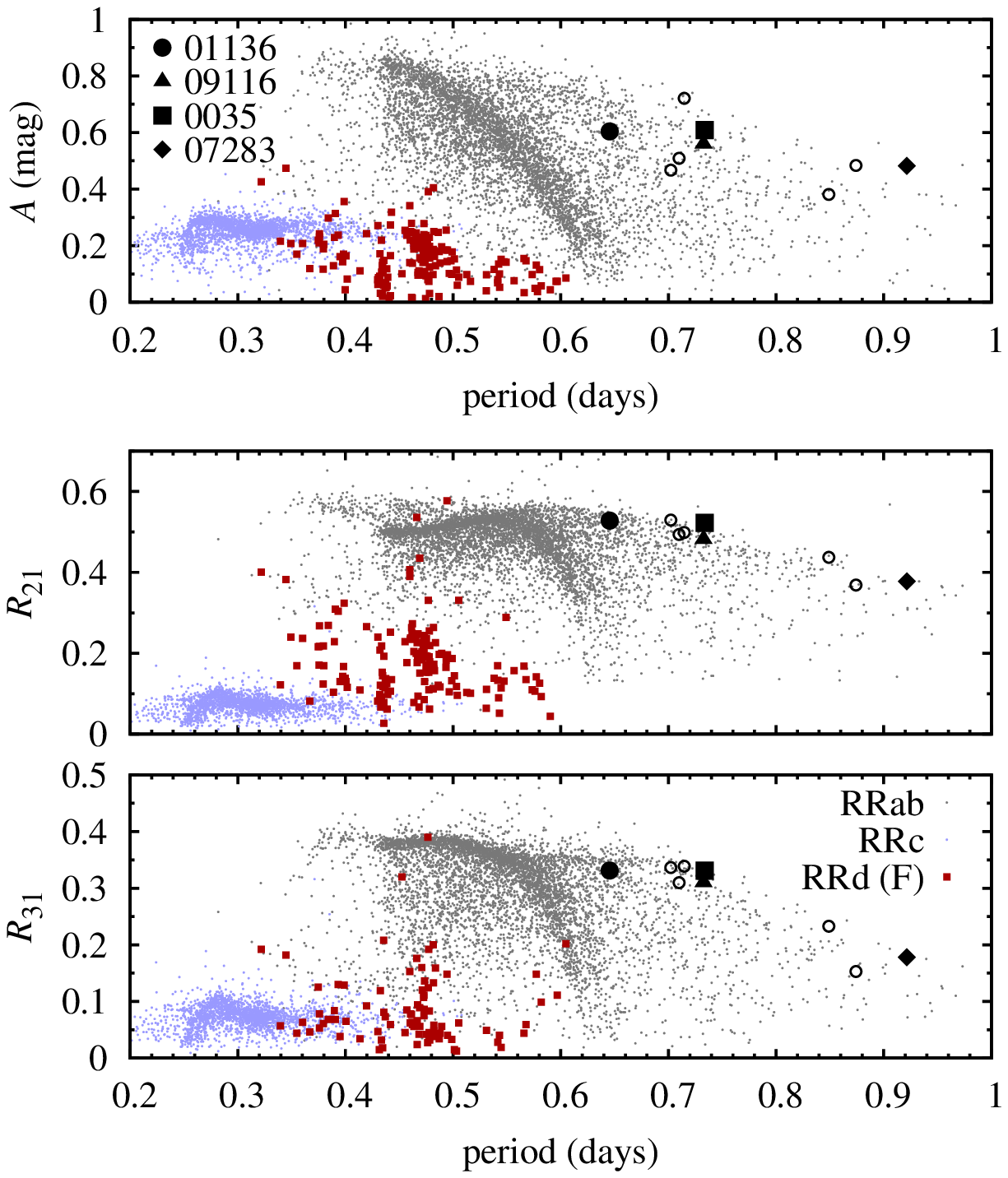}}\\
\resizebox{\hsize}{!}{\includegraphics{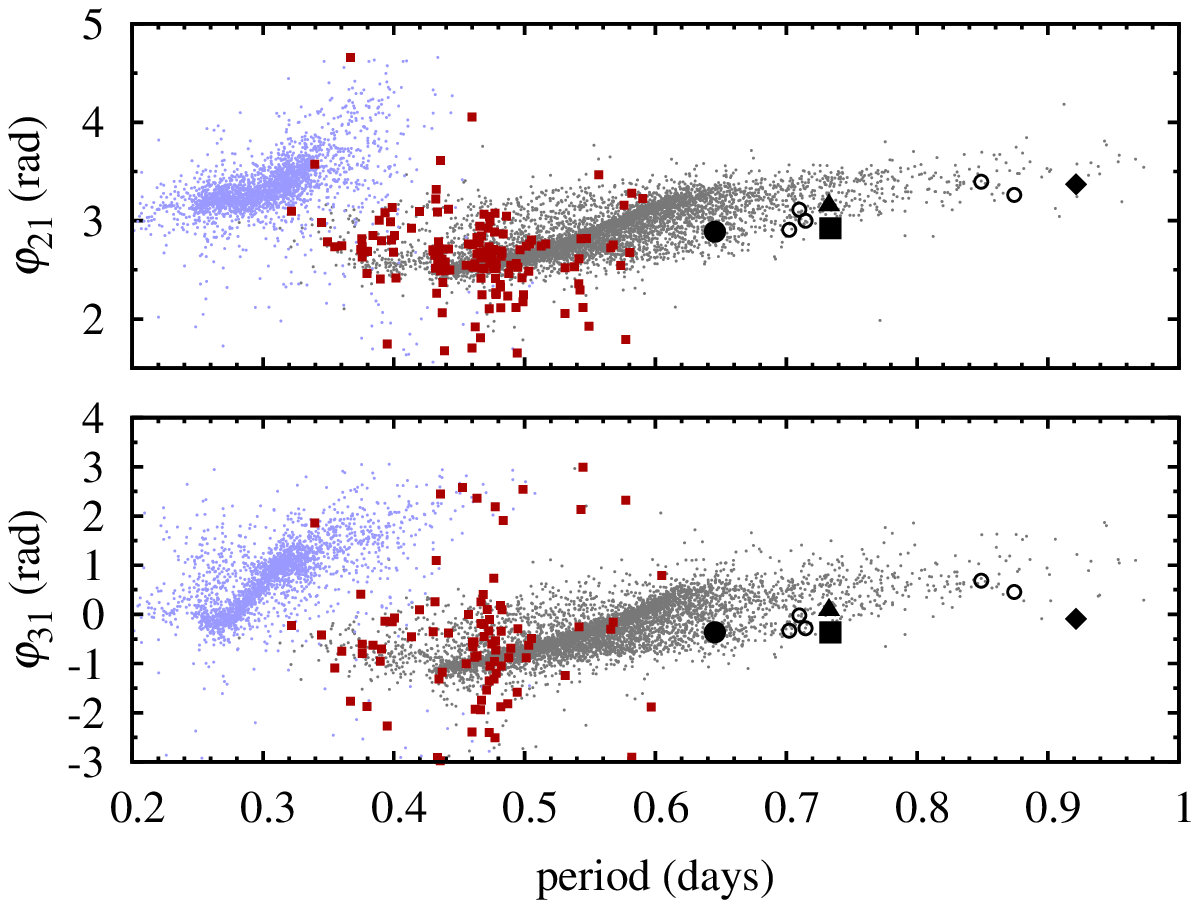}}
\caption{Fourier decomposition parameters for the extracted fundamental mode light curves of the four studied stars (large symbols) and for candidates (open circles). For comparison, Fourier parameters for RRab and RRc stars from the OGLE collection are plotted (for clarity, data for one half of the stars are plotted). In addition, the Fourier parameters of the fundamental mode in all 173 Galactic bulge OGLE RRd stars are included (small magenta squares).}
\label{fig:fdp}
\end{figure}

Before we discuss the nature of the additional variability, it is crucial to verify the nature of the dominant variability, attributed in the OGLE catalog to the radial fundamental mode. To this aim, we analyse the light curves corresponding to the dominant variability. These are obtained by subtracting from the data part of the light curve solution that contains all terms with frequencies different than $k\ff$. The resulting phased light curves are displayed in Fig.~\ref{fig:lc}. The corresponding low-order Fourier decomposition parameters \citep[$R_{k1}=A_k/A_1$, $\varphi_{k1}=\phi_k-k\phi_1$;][]{sl81} are plotted in Fig.~\ref{fig:fdp} and compared with the Fourier parameters for single periodic RRab and RRc stars from the OGLE collection \citep{ogleIV_rrl_blg}. 

The light curves (sorted by increasing period in Fig.~\ref{fig:lc}) are very characteristic for the long period RRab stars. This is fully confirmed with analysis of the Fourier decomposition parameters in Fig.~\ref{fig:fdp}. We note, that at a given period, the peak-to-peak amplitude and amplitude ratios are among the highest observed for RRab stars. The Fourier phases, on the other hand, are among the lowest. Still, such parameters are not exceptional; there are many RRab stars that share the same properties. The light curves leave no doubt; the dominant variability must correspond to radial fundamental mode.

It is instructive to compare the properties of the fundamental mode in the discussed stars with the properties of the fundamental mode in RRd stars. In Fig.~\ref{fig:fdp}, the small magenta squares represent the Fourier parameters computed for the fundamental mode for RRd stars of the OGLE-IV Galactic bulge collection \citep{ogleIV_rrl_blg}. We observe, that for the majority of RRd stars, the Fourier phases are among typical for the fundamental mode. However, the peak-to-peak amplitudes and amplitude ratios are significantly lower. This is well understood. In these stars two modes are simultaneously excited and in $\approx 82$ per cent the first overtone has the higher amplitude and dominates the pulsation. Since the two modes compete to saturate the pulsation instability, the fundamental mode's amplitude is lower than in the single-periodic RRab stars; its light curve is more sinusoidal and consequently the amplitude ratios are smaller. No such effect is visible for the four stars and five candidates discussed in this paper. As described above, the amplitudes and amplitude ratios are among the highest observed for RRab stars at a given period. This is because the additional variability in the discussed stars is always of very low amplitude, between 1.5 and 8.5 per cent of the fundamental mode amplitude (Tab.~\ref{tab:tab}). Except in 07283, additional variability may be described as a sine wave. Only in 07283 one harmonic, $2\fx$ is detected. Consequently, the additional variability is only a small perturbation in the otherwise fundamental mode pulsation of the four variables. 

Finally we note that light curves similar to that displayed in Fig.~\ref{fig:lc} are also observed in anomalous Cepheids pulsating in the fundamental mode. Pulsation periods of these stars are longer than 0.6 days. The largest known sample of these stars was reported in the OGLE Magellanic Clouds' photometry \citep{ogleIV_acep_mc}. The sample includes four anomalous Cepheids located in our Galaxy, in the foreground of the Magellanic Clouds. Comparison of the Fourier parameters of all anomalous Cepheids reported in \cite{ogleIV_acep_mc} with the Fourier parameters of the discussed stars, secures their RR~Lyr identification. Although the differences are not large, $R_{21}$, $\varphi_{21}$ and $\varphi_{31}$ are systematically smaller in the anomalous Cepheids.

\subsection{Notes on the individual stars}\label{ssec:notes}

\begin{figure}
\centering
\resizebox{\hsize}{!}{\includegraphics{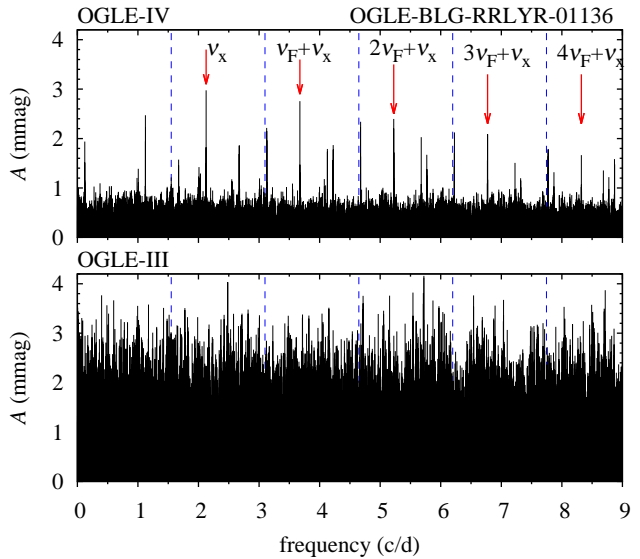}}
\caption{Top panel: frequency spectrum of the OGLE-IV data for OGLE-BLG-RRLYR-01136 after prewhitening with the fundamental mode and its harmonics (dashed lines). Additional periodicity and all detected combinations with the fundamental mode frequency are marked with arrows. Other significant peaks are their daily aliases. In the bottom panel we show the frequency spectrum for the same star, computed using the OGLE-III data. Additional periodicity is not detected because of a much larger noise level.}
\label{fig:fsp1136}
\end{figure}

{\it OGLE-BLG-RRLYR-01136}. In this star the additional periodicity has the lowest relative amplitude of only $1.5$\thinspace per cent (3 mmag) of the fundamental mode amplitude. Frequency spectrum after prewhitening with the fundamental mode and its harmonics is presented in the top panel of Fig.~\ref{fig:fsp1136} (OGLE-IV data). Six combination frequencies are clearly detected, all of the form $k\ff+\fx$, $k=1,\ldots, 6$. The additional mode is coherent, while for the fundamental mode a weak (S/N=3.9) residual power remains after the prewhitening. No additional significant periodicities are detected in the star. OGLE-III data are available, but the data sampling is much more sparse. Consequently, in the frequency spectrum the noise level is much higher. This is illustrated in the bottom panel of Fig.~\ref{fig:fsp1136}. Only the fundamental mode and its harmonics are detected; the possible signals of amplitude below 4 mmag are hidden in the noise.

\begin{figure}
\centering
\resizebox{\hsize}{!}{\includegraphics{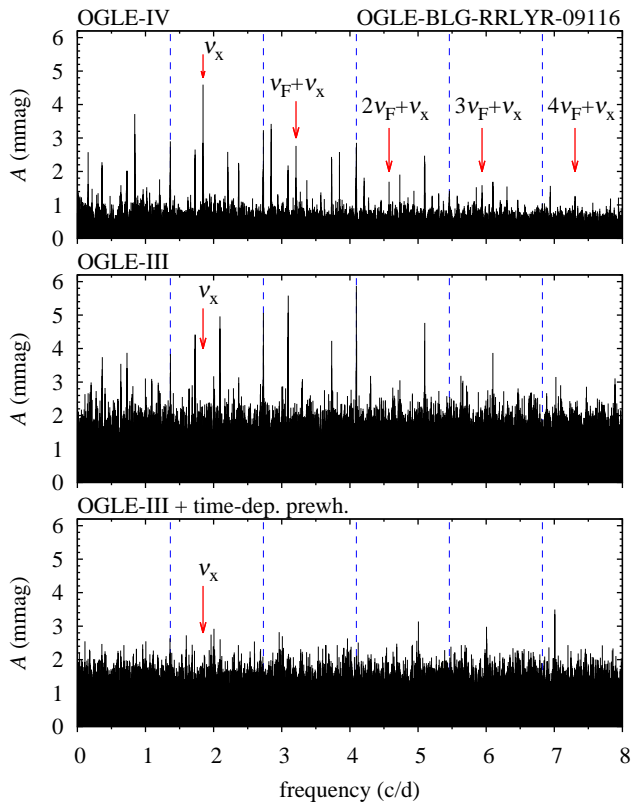}}
\caption{The same as Fig.~\ref{fig:fsp1136}, but for OGLE-BLG-RRLYR-09116. Note that fundamental mode is non-stationary as significant residual signal is detected at $k\ff$. In the OGLE-III data (middle panel) the noise level is larger. However, $\fx$, if it had the same amplitude as in the OGLE-IV data, should be easily detected. This is not the case, even after time-dependent prewhitening is applied to the data (bottom panel). The expected location of the additional variability is marked with arrow in the middle and bottom panels.}
\label{fig:fsp9116}
\end{figure}

 {\it OGLE-BLG-RRLYR-09116}. In the top panel of Fig.~\ref{fig:fsp9116}, we show the frequency spectrum for 09116 OGLE-IV data, after prewhitening with the fundamental mode and its harmonics. Additional periodicity is clearly present; a few combination frequencies may be identified as well. The analysis is hampered by the strong residual power at $k\ff$. It is a signature of a significant change of period of the fundamental mode, as we will demonstrate below. The OGLE-III data for the star are much more sparse, but their time-base is a bit longer. However, we cannot detect the additional periodicity in the prewhitened spectrum (the expected location is marked with an arrow in the middle panel of Fig.~\ref{fig:fsp9116}). If it had the same amplitude as in the OGLE-IV data, it should be detected. The frequency spectrum is dominated by the residual power at $k\ff$ and corresponding daily aliases. We can get rid of these signals using the time-dependent prewhitening. The result is illustrated in the bottom panel of Fig.~\ref{fig:fsp9116}. The noise level is a bit lowered, as compared to the middle panel; still, no sign of additional periodicity can be found. We conclude that the amplitude of the signal at $\fx$ changes over time; it must be below $\approx 2$\thinspace mmag during the seasons corresponding to OGLE-III observations. 

\begin{figure}
\centering
\resizebox{\hsize}{!}{\includegraphics{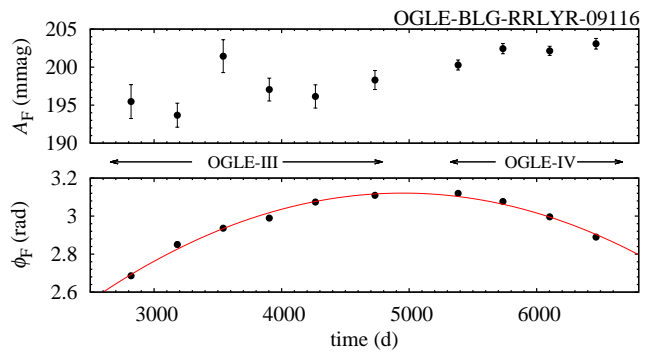}}
\caption{Time-dependent analysis of the amplitude and phase changes of the fundamental mode in 09116, on a season-to-season basis. Amplitude and phase variation are plotted in the top and in the bottom panels, respectively. Here
and in other figures time is expressed as $t={\rm HJD}-245\thinspace0000$\thinspace d.}
\label{fig:tdfd9116}
\end{figure}

The merged OGLE-III and OGLE-IV data can be used to conduct time-dependent Fourier analysis \citep{kbd87} for the fundamental mode. This is done on a season-to-season basis and results are presented in Fig.~\ref{fig:tdfd9116}. The amplitude of the fundamental mode vary, but the effect is not pronounced. The phase change, on the other hand, is pronounced. It is well described with the parabola, which corresponds to linear period increase at the rate of $1.1\times10^{-5}$\thinspace d per 1000\thinspace days.

 Unfortunately, OGLE-III data are too sparse to conduct any reliable time-dependent analysis of the possible variability of low-amplitude signal at $\fx$. Such analysis can be done using OGLE-IV data and indicates, that the amplitude of the additional mode was a bit lower during the first season of OGLE-IV observations (4\thinspace mmag). In the remaining three seasons the amplitude was stable at the level of 5\thinspace mmag. Some phase variation is also detected, but it is irregular and not as pronounced as for the fundamental mode.

\begin{figure}
\centering
\resizebox{\hsize}{!}{\includegraphics{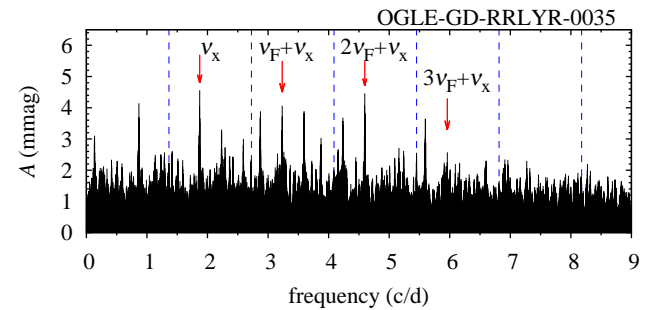}}
\caption{The same as Fig.~\ref{fig:fsp1136}, but for OGLE-GD-RRLYR-0035. Only OGLE-III data are available.}
\label{fig:fsp0035}
\end{figure}

{\it OGLE-GD-RRLYR-0035}. This star is located in the Galactic disc field; only OGLE-III data are available. After prewhitening with the fundamental mode and its harmonics, additional periodicity is clearly detected -- Fig.~\ref{fig:fsp0035}. Three combination frequencies are also detected, $\ff+\fx$, $2\ff+\fx$ and $3\ff+\fx$. In addition, another weak signal is detected at $\nu\approx 1.017$\thinspace c/d. Since no combinations with other detected signals are present, the signal is weak and its frequency is close to integer value, we do not consider it interesting. Both fundamental mode and additional periodicity are coherent.

\begin{figure}
\centering
\resizebox{\hsize}{!}{\includegraphics{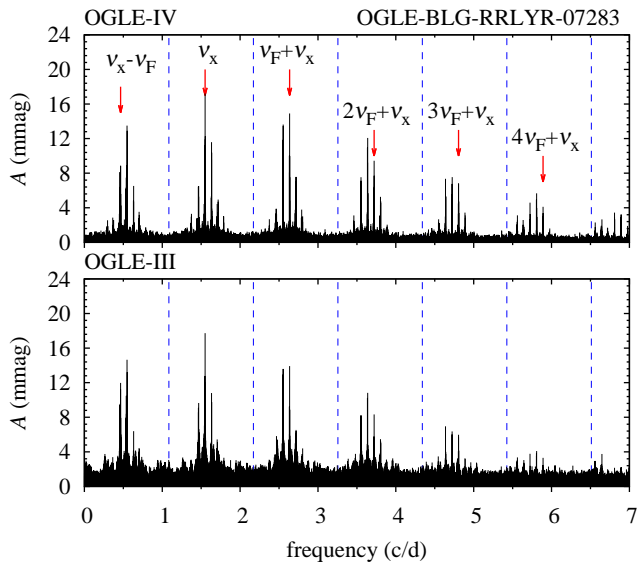}}
\caption{The same as Fig.~\ref{fig:fsp1136}, but for OGLE-BLG-RRLYR-07283. Many more combination frequencies are detected after prewhitening with the signals marked in the figure.}
\label{fig:fsp7283}
\end{figure}

\begin{figure}
\centering
\resizebox{\hsize}{!}{\includegraphics{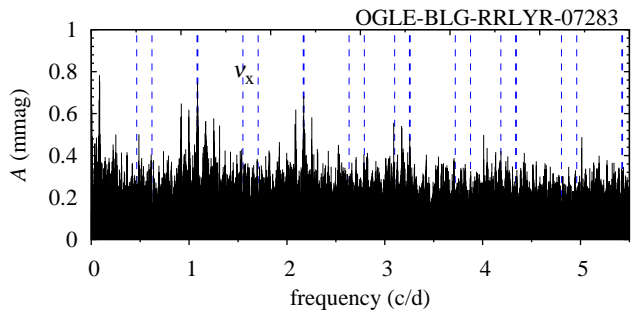}}
\caption{Frequency spectrum for 07283 OGLE-IV data after prewhitening with all significant frequencies (dashed lines; thick at $k\ff$).}
\label{fig:fsp7283_prewh}
\end{figure}

{\it OGLE-BLG-RRLYR-07283}. This star is the most interesting in our sample. It has the largest amplitude of the additional periodicity, $17$\thinspace mmag, which corresponds to more than $8$\thinspace per cent of the fundamental mode amplitude. Also, it is the brightest star in our sample (Tab.~\ref{tab:tab}) and its OGLE-IV data contain more than $8000$ data points over $4$ observing seasons (to be compared with $3092$ data points for the second, best-sampled case of 01136). Consequently, the noise level in the Fourier transform is very low and numerous additional signals may be detected in the frequency spectrum, as illustrated in Fig.~\ref{fig:fsp7283}. Altogether 34 significant frequencies are detected. Except fundamental mode and its 10 harmonics, we detect a signal at $\fx$ and its harmonic, $2\fx$, and 21 combination frequencies of the following forms: $k\ff+\fx$ ($k=1,\ldots, 11$), $\fx-\ff$, $k\ff-\fx$ ($k=2,\ldots, 9$) and $\ff+2\fx$. The frequency spectrum after prewhitening with the above frequencies is presented in Fig.~\ref{fig:fsp7283_prewh}. Note that the average signal level (noise) is only 0.14\thinspace mmag. Still, some significant signals are present, which are due to unresolved remnant power at $\ff$, $2\ff$ and $3\ff$ (plus daily aliases).

\begin{figure}
\centering
\resizebox{\hsize}{!}{\includegraphics{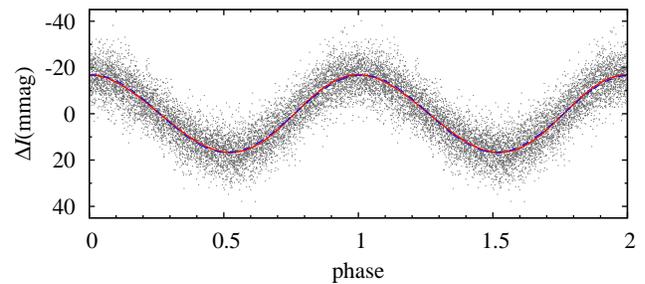}}
\caption{Phased light curve of the additional variability in 07283 with the second-order Fourier fit over-plotted (solid line). For comparison, a fitted sine wave is also plotted (dashed line).}
\label{fig:7283_lcX}
\end{figure}

07283 is the only star in our sample for which the harmonic of the additional variability could be detected. Consequently, the corresponding light variation is not a pure sine wave -- it is plotted in Fig.~\ref{fig:7283_lcX}. The difference with respect to sine wave (dashed line in Fig.~\ref{fig:7283_lcX}) is barely noticeable, however. The Fourier coefficient, $R_{21}$, of the secondary mode is only $0.040\pm 0.006$.

OGLE-II and OGLE-III data are available for the star (as a single data file); despite of their lower quality (2774 data points over 13 observing seasons) and larger noise level, the additional variability, as well as 12 combination frequencies, are clearly detected, see bottom panel of Fig.~\ref{fig:fsp7283}. 

The most intriguing property of 07283 is its period ratio, nearly exactly 7:10, $\pxpf=0.7000081(7)$ in the OGLE-IV data. The possible role played by the 7:10 resonance between the pulsation modes is discussed later in Section~\ref{sec:discussion}. The period ratio vary in time however; the average value over OGLE-II/III seasons (Tab.~\ref{tab:tab}) is $\pxpf=0.7000241(6)$ and the change with respect to the OGLE-IV data is almost entirely due to the increase of period of the additional variability. 

\begin{figure*}
\centering
\resizebox{\hsize}{!}{\includegraphics{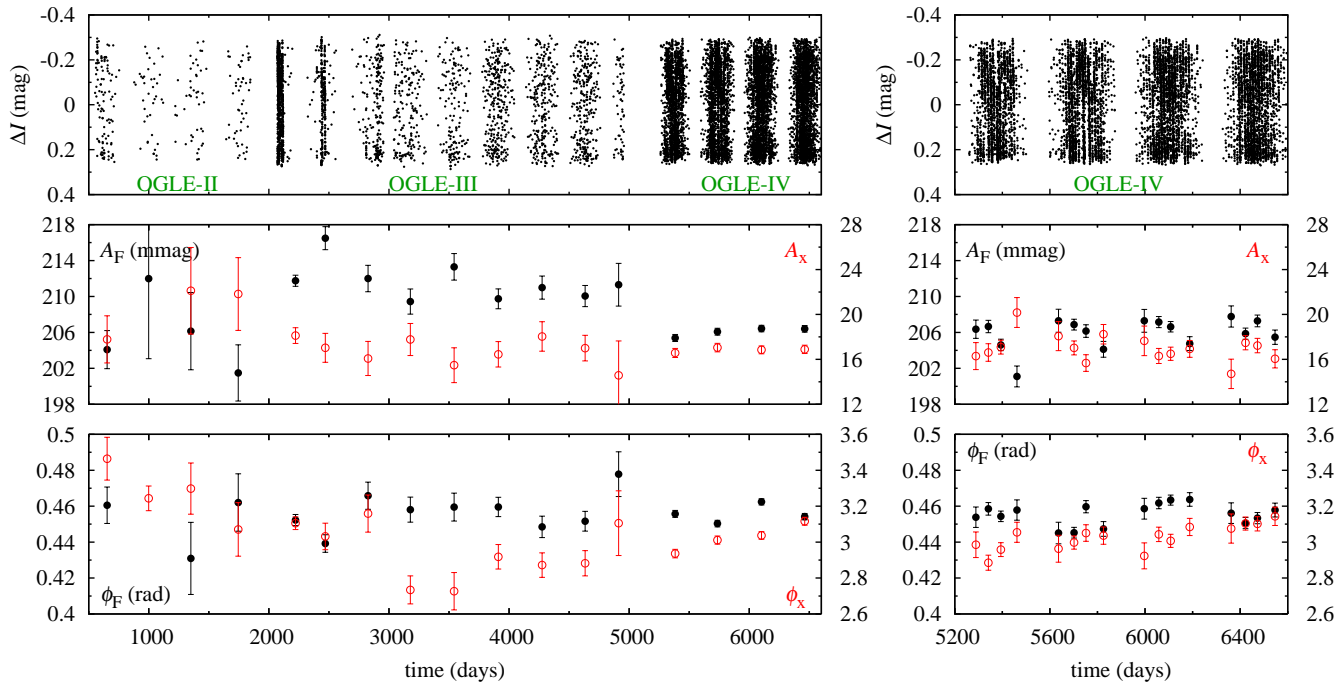}}
\caption{Time-dependent analysis of amplitude and phase changes of the two periodicities detected in 07283. In the top panel, the photometric data are plotted to visualise their quality. In the middle and bottom panels, amplitude and phase changes are plotted, respectively. Filled symbols correspond to the fundamental mode; its amplitude and phase may be read from the left-hand-side vertical axes. Open circles correspond to the additional variability; its amplitude and phase may be read from the right-hand-side vertical axes. In the left-hand panels, all data are analysed on a season-to-season basis, while the right-hand panels show the analysis of the top-quality OGLE-IV data with larger time resolution ($\approx 50$\thinspace d).}
\label{fig:7283_tdfd}
\end{figure*}

The long time span and good quality of the data, allow us to conduct a time-dependent analysis of amplitude and phase changes of the detected periodicities. The results are presented in Fig.~\ref{fig:7283_tdfd}. The photometric data are plotted in the top panel for a reference. Amplitude and phase changes are studied in the middle and bottom panels, respectively. We observe the following. (i) Amplitude of the fundamental mode clearly vary over the 17 observing seasons, but the variation may be largely of instrumental origin, as it seems to be correlated with the three phases of the OGLE experiment. The amplitude is, on average, lower during the four first seasons (OGLE-II). Then, it increases and remains more or less stable during the nine OGLE-III seasons. Finally, it decreases again for the last four OGLE-IV seasons. (ii) Considering the OGLE-IV seasons only (right-hand panels in Fig.~\ref{fig:7283_tdfd}), we observe that amplitude of the fundamental mode slightly decreases during each observing season. This suggest a possible periodic, low-amplitude modulation on a time-scale of one year. However, we do not find a clear signature of such modulation (equidistant triplet) in the frequency spectrum. The instrumental origin of this variation cannot be excluded. (iii) Some amplitude variation is also detected for the additional variability, but it is weak and no clear trends are visible. (iv) During the entire observations, the phase of the fundamental mode remains stable at the $3\sigma$ level. (v) For the last 10 observing seasons a trend of increasing phase of the additional variability is present (period change; see also Tab.~\ref{tab:tab}).

\subsection{Additional candidates for peculiar beat pulsation}\label{ssec:candidates}

We conducted a search for additional long-period, double-periodic stars similar to the four described above. The method was outlined in Sect.~\ref{sec:methods} and five detected candidates are reported in Tab.~\ref{tab:cand}. They are plotted in the Petersen diagram in Fig.~\ref{fig:pet} with open circles. In 08562 the period ratio is significantly higher than normally considered for RRd stars, while in 00984 it is lower (still significantly higher than in the extreme case of 07283). In all stars the additional periodicity was clearly detected with ${\rm S/N}>4.5$ (see the last column of Tab.~\ref{tab:cand}). However, only in 2 stars very weak signature of combination frequency ($\ff+\fx$) was present. Consequently, it is possible that we see a contamination, or just a random noise fluctuation. Nonetheless, when the light curve shape of the dominant fundamental mode variation is considered, these stars are very similar to the four discussed previously. This is demonstrated with the help of Fourier decomposition parameters, which are plotted in Fig.~\ref{fig:fdp} with open circles for the candidate stars. We observe that the peak-to-peak amplitudes and amplitude ratios are among the highest at the given period, while Fourier phases are a bit lower than in the majority of stars.

The above results indicate that the peculiar form of beat pulsation, we describe in this paper, may be more common among long-period RRab stars. Its detection clearly requires top-quality data.

\begin{table*}
\caption{Candidates for long-period RRab stars with additional periodicity in the $\pxpf\!\in\!(0.69,\,0.76)$ range, detected in the OGLE-IV Galactic bulge data. The consecutive columns contain: star's id, period of the fundamental mode, $\Pf$, period of the additional mode, $\Px$, period ratio, $\pxpf$, amplitude of the fundamental mode, $A_{\rm F}$, and amplitude ratio, $A_{\rm x}/A_{\rm F}$, remarks: S/N of the detection is given, `nsF/nsx' -- non-stationary fundamental mode/additional signal, `al' -- one-day alias is higher, `c' -- combination frequency present (S/N in the parenthesis), `ap' -- additional weak periodicity present.}
\label{tab:cand}
\begin{tabular}{lrrrrrr}
star's id & $\Pf$\thinspace (d) & $\Px$\thinspace (d) & $\pxpf$ & $A_{\rm F}$\thinspace (mag) & $A_{\rm x}/A_{\rm F}$ & remarks \\
\hline
OGLE-BLG-RRLYR-00984 &  0.7022298(2) & 0.50368(2) & 0.7173 & 0.1565(3) & 0.012 & S/N=4.6, ap\\ 
OGLE-BLG-RRLYR-08562 &  0.8742448(4) & 0.66310(2) & 0.7585 & 0.2061(4) & 0.015 & S/N=6.6, nsF\\ 
OGLE-BLG-RRLYR-09198 &  0.7147399(3) & 0.52257(1) & 0.7311 & 0.2536(7) & 0.020 & S/N=4.8, al, ap, c(3.9)\\ 
OGLE-BLG-RRLYR-09958 &  0.7098961(6) & 0.52130(1) & 0.7343 & 0.180(1)  & 0.046 & S/N=6.7, c(3.5)\\ 
OGLE-BLG-RRLYR-14334 &  0.8490460(7) & 0.61834(1) & 0.7283 & 0.1494(5) & 0.049 & S/N=8.9, nsx \\ 
\hline
\end{tabular}
\end{table*}

\section{Discussion}\label{sec:discussion}

\subsection{Nature of the additional variability}
The discussed stars seem to form a new group of peculiar double-periodic RR~Lyr pulsators of long period. The nature of the dominant variability seems well established -- the very characteristic light curve shape clearly identifies the radial fundamental mode. Additional variability is of shorter period, too short to be connected with rotation or binarity effects. 

As observed range of period ratios is very similar to that covered by RRd stars, the most natural explanation seems the excitation of additional pulsation mode. In principle, the additional low-amplitude variability may correspond to non-radial pulsation mode. \cite{vdk98} and \cite{dc99} show that low degree modes are excited in RR~Lyr models, preferentially in the vicinity of the radial modes. The frequency spectrum of such modes is very dense, however, and their growth rates are at least order of magnitude lower than of the neighbouring radial mode. Although the linear growth rates are not a good predictor of actual mode selection \citep[see e.g.][]{smolec14}, the possibility that isolated non-radial mode is excited in the discussed stars, instead of a strongly driven radial mode, seems unlikely. Below we show that the simultaneous excitation of two radial modes is the most plausible  solution for the discussed stars.

\begin{figure}
\centering
\resizebox{\hsize}{!}{\includegraphics{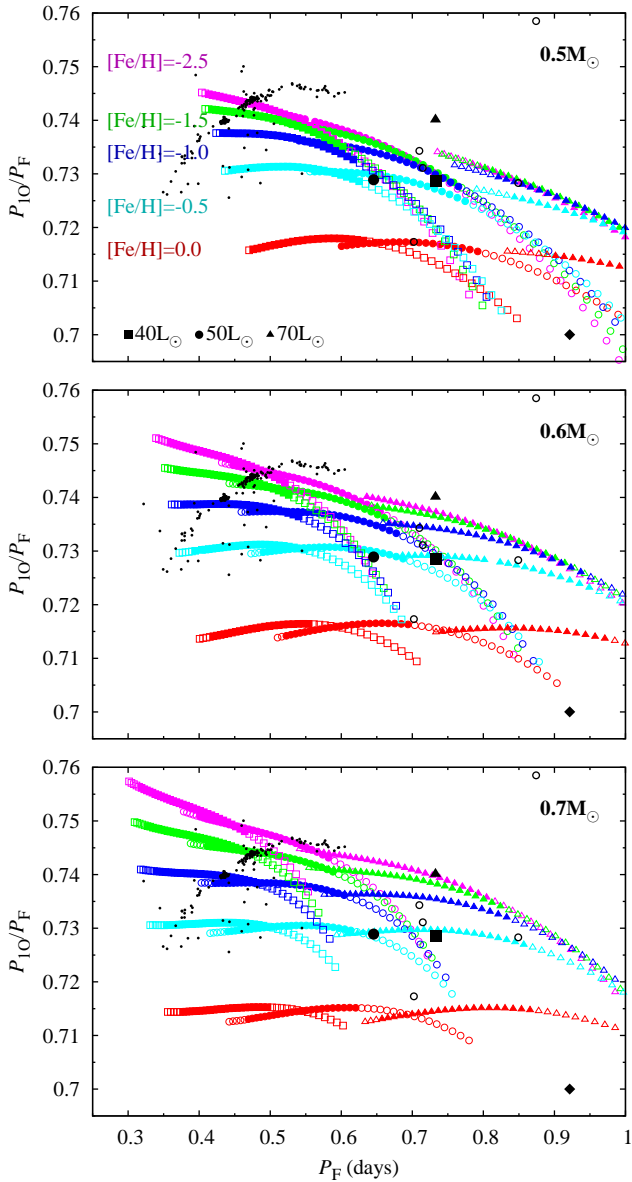}}
\caption{Period ratios for the Galactic bulge RRd stars (small dots) and long-period, double-periodic stars discussed in this paper (large symbols), confronted with the pulsation models. The three panels correspond to three different model masses, as indicated in the top right corners. Model sequences are plotted with different symbols, depending on the models' luminosity and with different colors, depending on the models' metallicity, as indicated in the top panel. Filled symbols correspond to the models in which fundamental mode and first overtone are simultaneously linearly unstable. Open symbols correspond to the models in which only one mode is linearly unstable.}
\label{fig:models1}
\end{figure}

In Fig.~\ref{fig:models1}, we confront the observed Petersen diagram with pulsation models of RR~Lyr-type stars. The models were computed with the pulsation codes of \cite{sm08a} that implement the \cite{kuhfuss} convection model. These are envelope models, consisting of $150$ mass shells, extending down to $2\times10^6$\thinspace K. The anchor zone, in which temperature is fixed to $11\,000$\thinspace K., is placed $40$ shells below the surface (it assures smooth variation of mode growth rates along a model sequence). All models use OPAL opacities \citep{opal} and adopt \cite{a09} solar mixture. Several model sequences were computed at fixed luminosity ($40$, $50$, $60$ or $70\LS$) and with different metallicities (${\rm [Fe/H]}=-2.5,\ldots,0.0$; step $0.5$\thinspace dex). Consecutive models along each sequence were computed with $25$\thinspace K step in effective temperature and cover the full extent of the instability strip. The three panels of Fig.~\ref{fig:models1} show period ratios, $P_{\rm 1O}/P_{\rm F}$, for pulsation models of different masses: $0.5\MS$ (top), $0.6\MS$ (middle) and $0.7\MS$ (bottom), and selected luminosities and metallicities. 

The necessary condition for the non-resonant, double-mode pulsation is simultaneous instability of the two pulsation modes. This condition is satisfied in the models plotted with filled symbols in Fig.~\ref{fig:models1}. The open symbols correspond to the models in which only one mode is linearly unstable (first overtone at shorter periods and fundamental mode at longer periods). Inspection of Fig.~\ref{fig:models1} reveals: 
(i) For nearly all stars discussed in this paper, we can find the models in which both radial modes are simultaneously unstable; consequently, non-resonant fundamental and first overtone double-mode pulsation seems a plausible explanation for the observed double-periodic variability. 
(ii) We can reproduce the location of 09116 in the Petersen diagram only assuming large mass ($0.7\MS$) and very low metallicity (${\rm [Fe/H]}=-2.5$; bottom panel of Fig.~\ref{fig:models1}). 
(iii) Position of 01136 can be matched with models of all considered masses, however, only with large metallicity, ${\rm [Fe/H]}\approx-0.5$, the two modes are simultaneously unstable. 
(iv) Position of 0035 is also well reproduced with  ${\rm [Fe/H]}\approx-0.5$ and all considered masses. For the lowest mass, $0.5\MS$, the two radial modes are unstable also at significantly lower metallicities. 
(v) 07283, with its long period, is the most challenging star to model. Its period ratio can be matched with the models, however, first overtone is always stable. In fact, the models predict that 07283 is at the red edge of the fundamental mode instability strip, point that its luminosity should be around $50\LS$, and mass below $0.6\MS$, as only then the fundamental mode is unstable. In principle, position of 07283 can be fit assuming any metallicity. 
(vi) Plausible models can be found for all candidates (Sect.~\ref{ssec:candidates}), except 08562; its high period ratio cannot be reproduced by any RR~Lyr-type models assuming excitation of two radial modes. Still, this is only a candidate star. The additional variability may be just due to contamination. 

Clearly, the location of 07283 in the Petersen diagram is the most difficult to explain assuming it is an extreme RRd star. Not surprisingly, at such long period ($P_{\rm 1O}\approx 0.645$\thinspace d), first overtone is firmly stable. The mode stability depends on the convection parameters in the model. In the extreme case of purely radiative models, the red edge is not present at all. In principle, we can produce the unstable 1O e.g. through a significant decrease of mixing-length parameter or eddy-viscosity parameter; but such models are unrealistic, then. Note that the longest period of RRc star in the OGLE Galactic bulge collection is $\approx 0.54$\thinspace d (see also Fig.~\ref{fig:fdp}). The obvious circumvention is the resonant excitation of the first overtone. In the light of a very peculiar period ratio in this star, nearly exactly 0.7, the 7:10 resonance seems the solution. In such case, first overtone is  linearly dumped, and excited to finite amplitude by resonant coupling with the fundamental mode. The proposed resonance is of extremely high order, however; all other resonances, that are known to have crucial effects on the pulsation of classical pulsators, are of lower order. These are the 2:1 resonances causing e.g. the Cepheid bump progression \citep[e.g.][]{bmk90} or half-integer resonances causing the period doubling phenomena \citep{mb90}. Of these, the 9:2 resonance, that is suspected of causing period doubling in Blazhko RR~Lyr stars \citep{kms11}, is of the highest order. Also, the exact phase synchronisation is not observed in 07283; it is only in the near-resonant condition. Consequently, we judge the possible role of the 7:10 resonance in the excitation of the first overtone as unlikely. 

The other solution might be a different nature of 07283. Recently, a new group of pulsating variable stars was identified, light curves of which may mimic the RR~Lyrae light curves \citep{bepnat,bep}. These are the so-called binary evolution pulsators (BEPs), with masses much lower than in RR~Lyr stars. The first, and so far the only firm member of the group, OGLE-BLG-RRLYR-02792, has a mass of only $0.26\MS$. We have computed a grid of pulsation models assuming much lower model masses ($0.2-0.4\MS$) to check whether the location of 07283 in the Petersen diagram can be reproduced then. Indeed, it can, but exactly the same problem as in the case of RR~Lyr-type models arises; 1O is firmly stable. 

At the moment we do not have a satisfactory explanation for the nature of additional variability in 07283. As dominant variability clearly corresponds to the radial fundamental mode, and the observed period ratio can be well modelled assuming the concomitant excitation of the radial first overtone, the RRd nature seems most likely. The excitation mechanism for the first overtone remains unknown, however.

Simultaneous instability of the two radial modes is the necessary condition for the double-mode pulsation. Still, the star can {\it choose} to pulsate in one radial mode, only. These are the non-linear effects that determine the actual pulsation form. The mechanism of mode selection is a very difficult, non-linear and not well understood problem \citep[for a review see][]{smolec14}. It is well visible in Fig.~\ref{fig:models1} that simultaneous instability of the fundamental and first overtone modes occurs over wide areas in the Petersen diagram. Still, the majority of RRd stars fall along a tight, well defined progression, location of which is determined by  the non-linear effects. The four long-period, double-periodic stars, and candidates discussed in Section~\ref{ssec:candidates}, do not belong to this progression. In fact, they may belong to a different group or different population of RR~Lyrae stars.

\subsection{Population membership of the extreme RRd stars}

\begin{figure}
\centering
\resizebox{\hsize}{!}{\includegraphics{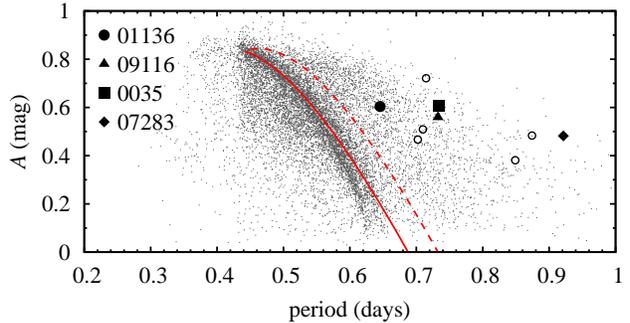}}
\caption{Period amplitude diagram for Galactic bulge RRab stars. The solid line marks the loci of the OoI group, the dashed line separates the OoI and OoII groups.}
\label{fig:pa}
\end{figure}

The population studies of RR~Lyrae stars are based on the period-amplitude (Bailey) diagram and on metallicity determinations. For globular clusters, the so-called Oosterhoff groups are defined based on the location of the cluster in the mean period -- mean metallicity diagram \citep[for a review, see e.g.][]{catelan09}. Members of the Oosterhoff group I (OoI) have, on average, shorter periods and higher metal abundances, while members of the Oosterhoff group II (OoII) have, on average, longer pulsation periods and lower metal abundances. The Oosterhoff group III (OoIII) was also identified, with the longest pulsation periods and the highest, close to solar metallicities \citep{OoIII}. The Oosterhoff affiliation of individual star may be determined by its location in the Bailey diagram. In this diagram, the Oo groups form ridges that progressively shift towards longer periods from OoI to OoIII. The Oosterhoff analysis may also be applied to field RR~Lyrae stars, although separation between the Oo groups is not well defined then \citep[e.g.][]{kunder,dszczyg}. The majority of the OGLE Galactic bulge RR~Lyrae stars are members of the OoI, which we illustrate in the Bailey diagram in Fig.~\ref{fig:pa}. The OoI ridge, plotted with the solid line, was determined following the procedure outlined in \cite{zorotovic}. The dashed line separates the OoI and OoII groups adopting the \cite{miceli} criterion. Details of these procedures will be published elsewhere (Prudil, in prep.). To the right of the dashed line the OoII group extends. At the longest periods, members of the OoIII group may be expected. All stars discussed in this paper are thus members of OoII (or OoIII), while the majority of RRd stars are OoI. 

As mentioned above, the Oo membership is correlated with metallicity. Unfortunately, we lack spectroscopic metallicity determination for the discussed stars. The pulsation models allowed us to put some rough constraints on the metallicities, e.g. 09116 should be metal poor (${\rm [Fe/H]}\!\approx\!-2.5$), while for the 01136 high metallicity is preferred (${\rm [Fe/H]}\!\approx\!-0.5$). For 07283 and 0035 a range of metallicities is in principle allowed. The metallicity can also be determined using the photometric method, based on the light curve shape \citep{jk96}. The method was calibrated for the $I$-band in \cite{sm05} and in Tab.~\ref{tab:feh} we report the results obtained with their eq.~2 [${\rm [Fe/H]}=f(P,\,\varphi_{31})$]. The metallicity was computed using the $\varphi_{31}$ of the fundamental mode light curves (Fig.~\ref{fig:lc}). The calculated values are typical for RR~Lyr stars. We note that for 09116 the photometric metallicity is higher than predicted by the pulsation models (${\rm [Fe/H]}\approx-1.5$ vs. ${\rm [Fe/H]}\approx-2.5$), while for 01136 it is lower (${\rm [Fe/H]}\approx-1.4$ vs. ${\rm [Fe/H]}\approx-0.5$). The differences are of order of $1$\thinspace dex and are significant. A word of caution is needed, however. The majority of stars in the calibration sample of \cite{sm05} are short period field stars. The longest pulsation period in the calibrating sample was $\approx 0.7$\thinspace days. As noted by \cite{dszczyg}, the OoII stars might follow a different metallicity-period-phase relation than OoI stars. Thus, the computed photometric metallicities may be a subject to a systematic error.

\begin{table}
\centering
\caption{Metallicity estimates for the discussed stars using the photometric method. }
\label{tab:feh}
\begin{tabular}{lr}
star & [Fe/H] \\
\hline
OGLE-BLG-RRLYR-01136 & $-1.43$ \\
OGLE-BLG-RRLYR-09116 & $-1.49$ \\
OGLE-GD-RRLYR-0035   & $-1.86$ \\
OGLE-BLG-RRLYR-07283 & $-2.56$ \\
\hline
\end{tabular}
\end{table}

\section{Summary}\label{sec:summary}

We report on the discovery of four, long period, double-periodic stars in the OGLE observations of the Galactic bulge and Galactic disc. The original classification of these stars in the OGLE collection is RRab. Indeed, the dominant variability corresponds to the radial fundamental mode, as a very characteristic light curve shape indicates. The additional, shorter period variability is of low amplitude, up to $8.5$\thinspace per cent of the fundamental mode amplitude (typically $2-5$\thinspace mmag, nearly $18$\thinspace mmag in the extreme case of 07283). The additional variability may be described with a sine wave, except for 07283, in which one harmonic is present. The additional variability most likely corresponds to the radial first overtone. The observed period ratios ($P/\Pf\!\in\!(0.7,\,0.74)$) can be well reproduced with the pulsation models, in which both radial modes are linearly unstable, except the extreme case of the longest period star (07283; $P_{\rm F}\approx0.92$\thinspace d), in which the first overtone is firmly stable. In this case the excitation mechanism for the first overtone remains unknown. Still, the interpretation of all stars as extreme RRd pulsators seems the most plausible. Five additional candidates, showing similar form of long-period, beat pulsation, were also identified. In all these stars the light curve shapes of the dominant fundamental mode, when described with the help of the Fourier decomposition parameters, share a common characteristics. The amplitudes and amplitude ratios are among the highest observed in single-periodic RRab stars at a given pulsation period, while the Fourier phases are among the lowest observed at a given pulsation period. 

As compared to the majority of RRd stars observed e.g. in the Galactic bulge, the extreme double-periodic stars  do not follow the same progression in the Petersen diagram. Their pulsation periods are much longer; fundamental mode strongly dominates the pulsation and first overtone is only a small perturbation. The different nature of the discussed stars is also apparent in the Bailey diagram.  In the Oosterhoff classification, they are either members of OoII, or OoIII groups. The origin of the Oosterhoff phenomenon still lacks satisfactory explanation; its studies for the field stars are scarce. There is no obvious connection between the peculiar RRd pulsation and Oo classification; the ordinary RRd stars populate both the OoI and OoII globular clusters.

\section*{Acknowledgements}
This research is supported by the National Science Centre, Poland, through grants DEC-2012/05/B/ST9/03932 and DEC-2015/17/B/ST9/03421. MS acknowledges the support of the postdoctoral fellowship programme of the Hungarian Academy of Sciences at the Konkoly Observatory as a host institution, and the support by Hungarian National Research, Development and Innovation Office -- NKFIH K-115709. KB is supported by NCN grants DEC-2012/07/N/ST9/04172 and DEC-2015/16/T/ST9/00174. We are grateful to Pawel Moskalik and Wojtek Dziembowski for commenting the manuscript and fruitful discussions.


\bsp	
\label{lastpage}

\begin{thebibliography}{99}
\bibitem[\protect\citeauthoryear{Asplund et al.}{2009}]{a09} Asplund M., Grevesse N., Sauval A. J., Scott P., 2009, ARA\&A, 47, 481
\bibitem[\protect\citeauthoryear{Benk\H{o} et al.}{2010}]{benko10} Benk\H{o} J.M., et al., 2010, MNRAS, 409, 1585
\bibitem[\protect\citeauthoryear{Benk\H{o} et al.}{2014}]{benko14} Benk\H{o} J.M., et al., 2014, ApJ Suppl. Ser., MNRAS, 213, 31
\bibitem[\protect\citeauthoryear{Buchler, Moskalik \& Kov\'acs}{1990}]{bmk90} Buchler J.R., Moskalik P., Kov\'acs G., 1990, ApJ, 351, 617
\bibitem[\protect\citeauthoryear{Catelan}{2009}]{catelan09} Catelan M., 2009, Astrophys. Space Sci., 320, 261
\bibitem[\protect\citeauthoryear{Dziembowski}{2016}]{wd16} Dziembowski W., 2016, Comm. Konkoly Obs., vol. 105, 23
\bibitem[\protect\citeauthoryear{Dziembowski \& Cassisi}{1999}]{dc99} Dziembowski W.A., Cassisi S., 1999, Acta Astron., 49, 371
\bibitem[\protect\citeauthoryear{Gruberbauer et al.}{2007}]{aqleo} Gruberbauer M., Kolenberg K., Rowe J. et al., 2007, MNRAS, 379, 1498
\bibitem[\protect\citeauthoryear{Iglesias \& Rogers}{1996}]{opal} Iglesias C.A., Rogers F.J., 1996, ApJ, 464, 943
\bibitem[\protect\citeauthoryear{Jurcsik \& Kov\'acs}{1996}]{jk96} Jurcsik J., Kov\'acs G., 1996, A\&A, 312, 111
\bibitem[\protect\citeauthoryear{Jurcsik et al.}{2014}]{jurcsikM3} Jurcsik J., et al., 2014, ApJ 797, L3
\bibitem[\protect\citeauthoryear{Koll\'ath, Moln\'ar \& Szab\'o}{2011}]{kms11}  Koll\'ath Z., Moln\'ar L., Szab\'o R., 2011, MNRAS, 414, 1111
\bibitem[\protect\citeauthoryear{Kov\'acs, Buchler \& Davis}{1987}]{kbd87}  Kov\'acs G., Buchler J.R., Davis C.G., 1987, ApJ, 319, 247
\bibitem[\protect\citeauthoryear{Kuhfu\ss{}}{1986}]{kuhfuss} Kuhfu\ss{} R., 1986, A\&A, 160, 116
\bibitem[\protect\citeauthoryear{Kunder \& Chaboyer}{2009}]{kunder} Kunder A., Chaboyer B., 2009, AJ, 138, 1284
\bibitem[\protect\citeauthoryear{Miceli et al.}{2008}]{miceli} Miceli A., et al., 2008, ApJ, 678, 865
\bibitem[\protect\citeauthoryear{Moln\'ar}{2016}]{molnarRRL15} Moln\'ar L., 2016, Comm. Konkoly Obs., vol. 105, 11
\bibitem[\protect\citeauthoryear{Moln\'ar et al.}{2012}]{molnar12} Moln\'ar L., Koll\'ath Z., Szab\'o R., Bryson S., Kolenberg K., Mullally F., Thompson S.E., 2012, ApJ, 757, L13
\bibitem[\protect\citeauthoryear{Moskalik \& Buchler}{1990}]{mb90} Moskalik P., Bucher J.R., 1990, ApJ, 355, 590
\bibitem[\protect\citeauthoryear{Moskalik et al.}{2015}]{pamsm15} Moskalik P., Smolec R., Kolenberg K. et al., 2015, MNRAS, 447, 2348
\bibitem[\protect\citeauthoryear{Netzel \& Smolec}{2016}]{nspta} Netzel H., Smolec R., 2016, Proceedings of the Polish Astron. Soc., in press, preprint arXiv:1603:05389
\bibitem[\protect\citeauthoryear{Netzel, Smolec \& Moskalik}{2015a}]{netzel1} Netzel H., Smolec R., Moskalik P., 2015a, MNRAS, 447, 1173
\bibitem[\protect\citeauthoryear{Netzel, Smolec \& Dziembowski}{2015}]{netzel2} Netzel H., Smolec R., Dziembowski W., 2015, MNRAS, 451, L25
\bibitem[\protect\citeauthoryear{Netzel, Smolec \& Moskalik}{2015b}]{netzel3} Netzel H., Smolec R., Moskalik P., 2015b, MNRAS, 453, 2022
\bibitem[\protect\citeauthoryear{Pietrukowicz et al.}{2013}]{pietruk} Pietrukowicz P., et al., 2013, Acta Astron., 63, 379
\bibitem[\protect\citeauthoryear{Pietrzy\'nski et al.}{2012}]{bepnat} Pietrzy\'nski et al., 2012, Nature, 484, 75
\bibitem[\protect\citeauthoryear{Pritzl et al.}{2000}]{OoIII} Pritzl B., Smith H.A., Catelan M., Sweigart A.V., 2000, ApJ, 530, L41
\bibitem[\protect\citeauthoryear{Simon \& Lee}{1981}]{sl81} Simon N.R., Lee A.S., 1981, ApJ, 248, 291
\bibitem[\protect\citeauthoryear{Skowron et al.}{2016}]{skowron} Skowron J., et al., 2016, Acta Astron., 66, 1
\bibitem[\protect\citeauthoryear{Smolec}{2005}]{sm05} Smolec R., 2005, Acta Astron., 55, 59
\bibitem[\protect\citeauthoryear{Smolec}{2014}]{smolec14} Smolec R., 2014, IAUS, 301, 265
\bibitem[\protect\citeauthoryear{Smolec}{2016}]{sm15bl} Smolec R., 2016, Proceedings of the Polish Astron. Soc., in press, preprint arXiv:1603:01252
\bibitem[\protect\citeauthoryear{Smolec \& Moskalik}{2008}]{sm08a} Smolec R., Moskalik P., 2008, Acta Astron., 58, 193
\bibitem[\protect\citeauthoryear{Smolec et al.}{2013}]{bep} Smolec R., et al., 2013, MNRAS, 428, 3034
\bibitem[\protect\citeauthoryear{Smolec et al.}{2015a}]{rs15a} Smolec R., Soszy\'nski I., Udalski A., et al., 2015a, MNRAS, 447, 3756
\bibitem[\protect\citeauthoryear{Smolec et al.}{2015b}]{rs15b} Smolec R., Soszy\'nski I., Udalski A., et al., 2015b, MNRAS, 447, 3873
\bibitem[\protect\citeauthoryear{Soszy\'nski et al.}{2011}]{ogleIII_rrl_blg} Soszy\'nski I., et al., 2011, Acta Astron., 61, 1
\bibitem[\protect\citeauthoryear{Soszy\'nski et al.}{2014}]{ogleIV_rrl_blg} Soszy\'nski I., et al., 2014, Acta Astron., 64, 177
\bibitem[\protect\citeauthoryear{Soszy\'nski et al.}{2015}]{ogleIV_acep_mc} Soszy\'nski I., et al., 2015, Acta Astron., 65, 233
\bibitem[\protect\citeauthoryear{Soszy\'nski et al.}{2016}]{ogleIV_rrl_mc} Soszy\'nski I., et al., 2016, Acta Astron., submitted, e-print arXiv:1606.02727
\bibitem[\protect\citeauthoryear{Szab\'o}{2014}]{szabobl} Szab\'o R., 2014, in Guzik J. A., Chaplin W. J., Handler G., Pigulski A., eds, Proc. IAU Symp. 301, Precision Asteroseismology. Cambridge Univ. Press, Cambridge, p. 241
\bibitem[\protect\citeauthoryear{Szab\'o et al.}{2014}]{szabo_corot} Szab\'o R., Benk\H{o} J.M., Papar\'o M., 2014, A\&A, 570, A100
\bibitem[\protect\citeauthoryear{Szczygie\l{} et al.}{2009}]{dszczyg} Szczygie\l{} D.M., Pojma\'nski G., Pilecki B., 2009, Acta Astron., 59, 137
\bibitem[\protect\citeauthoryear{Udalski et al.}{2008}]{ogleIII} Udalski A., Szyma\'nski M.K., Soszy\'nski I., Poleski R., 2008, Acta Astron., 58, 69
\bibitem[\protect\citeauthoryear{Udalski, Szyma\'nski \& Szyma\'nski}{2015}]{ogleIV} Udalski A., Szyma\'nski M.K., Szyma\'nski G., 2015, Acta Astron., 65, 1
\bibitem[\protect\citeauthoryear{Van Hoolst, Dziembowski \& Kawaler}{1998}]{vdk98} Van Hoolst T., Dziembowski W.A., Kawaler S.D., 1998, MNRAS, 297, 536
\bibitem[\protect\citeauthoryear{Zorotovic et al.}{2010}]{zorotovic} Zorotovic M., et al., 2010, AJ, 139, 357


\end{thebibliography}
\end{document}